\documentclass{elsart}

\usepackage[english]{babel}
\usepackage[latin2]{inputenc}
\usepackage[T1]{fontenc}

\usepackage{pslatex}
\usepackage{fancyhdr}
\usepackage{ae,aecompl}
\usepackage{pst-eps}
\usepackage{graphics}
\usepackage{graphicx}
\usepackage{amsmath}
\usepackage{amssymb}
\usepackage{pifont}
\usepackage[dvips]{color}

\begin{document}

\begin{frontmatter}
\title{The effect of disorder on the hierarchical modularity in
  complex systems} 
\author[bme]{D. Nagy},
\author[bme]{G. Tibély},
\author[bme,lce]{J. Kertész}
\address[bme]{Department of Theoretical Physics, Budapest University of Technology and Economics, Budafoki út 8, H-1111 Budapest, Hungary}
\address[lce]{Laboratory of Computational Engineering, Helsinki University of Technology, P.O.Box 9400, FIN-02015 HUT, Finland}
\begin{abstract}

Hierarchically modular systems show a sequence of scale separations in
some functionality or property in addition to their hierarchical
topology. Starting from regular, deterministic objects like the
Vicsek snowflake or the deterministic scale free network by Ravasz
et al., we first characterize the hierarchical modularity by the
periodicity of some properties on a logarithmic scale indicating
separation of scales. Then we introduce randomness by keeping the
scale freeness and other important characteristics of the objects and
monitor the changes in the modularity. In the presented examples
sufficient amount of randomness destroys hierarchical modularity. Our
findings suggest that the experimentally observed hierarchical
modularity in systems with algebraically decaying clustering
coefficients indicates a limited level of randomness.

\end{abstract}
\end{frontmatter}

\section{INTRODUCTION}
\thispagestyle{plain}
During the last few years a great number of discoveries have formed
our view of complex networks. It turned out that the scale-free
property is an ubiquitous feature of most real networks, like the
World Wide Web, metabolic networks, and collaboration networks, etc.
\cite{Barabási,Newman}. The hierarchical topology of most of
these networks has also been studied \cite{RavaszBarabási,hierstruct}. 
According to recent studies \cite{Ravasz,Oltvai}, a 
hierarchically modular organization lies beneath this topology: 'there 
are many highly integrated small modules which group into a few larger 
modules, which in turn can be integrated into even larger modules' 
\cite{Ravasz}. The so-called \emph{hierarchical network model} was 
introduced \cite{hierstruct}, which is a simple illustration of the
idea above. However, the modularity of real networks is still an open 
question. The problem of the unambiguous identification of modules or 
communities of a system at different hierarchical levels has not been 
solved ultimately, though important advances have been achieved 
\cite{community}\footnote{In the paper by Clauset et al. 
\cite{community} a quantity called modularity is defined based on 
already identified modules. We want to avoid this a priori
identification (which is often a hard task) and we use the term
modularity in the every-day sense for systems consisting of well
identifiable and separable composite units.}. The 
\emph{average-linkage hierarchical clustering algorithm} \cite{Eisen} 
groups the points of a network according to the \emph{topological
overlap} between them \cite{Ravasz}. Modularity is then attributed to
the clustered structure of the overlaps and hierarchical modularity is
obtained if the observed clusters can be interpreted such, i.e. if the 
clusters are hierarchically nested. The advantage of this method is 
that it avoids the delicate problem of a priori identification of 
modules. 

In Ref. \cite {Ravasz} metabolic networks were studied and the interesting
conclusion was drawn that complex networks show hierarchical
modularity if they are characterized by a clustering coefficient
$C(k)$ decaying with a power law as a function of the degree $k$ of the
nodes (often with an exponent close to unity). The clustering
coefficient is a measure of the inter-connectedness of the
neighbors of a particular node \cite {Barabási,Newman,Dorogov} and
it is clearly related to the community structure. In Ref. \cite {Ravasz}
this concept was nicely illustrated by a regular network and used to
analyse the experimental data obtained for metabolic networks.

In the present work we address the question of the effect of
randomness on the hierarchical modularity. In order to do so we 
introduce a measure of modularity in scale free (hierarchical) systems
without using identification of modules. We apply this concept to 
randomly rearranged regular fractals and networks.

\section{MODULARITY OF FRACTALS}
\label{sec:Fractals}

\subsection{The Vicsek Snowflake}
Understanding the modularity of networks is rather difficult, because
it is hidden in the network's topology. Therefore, first we show the
effect of modularity in the case of fractals. We investigated the
modularity of \emph{randomly rearranged} Vicsek snowflakes embedded 
into two dimensions.\newline Our initial object was the well known 
fractal shown on the left of Fig. \ref{fig:otherps1}, which has 
regular self-similarity \cite{Snowflake}. At the highest hierarchical 
level, it consists of five well-separated blocks, and each of these
blocks contains five smaller blocks in the same fashion, and so on. In 
our case, this hierarchical structure has a finite resolution, so it 
repeats itself until reaching an elementary block size or lower 
cutoff. The fractal's dimension is given by the 
formula of $D = \ln 5/\ln 3 \approx 1.465$.  Obviously, for the 
regular fractal, the blocks (which have a finer structure not resolved 
at this level) can be considered as modules at each level of the 
hierarchy. As these modules are intact until the next, finer level of 
the hierarchy is reached, there is a clear sequence of separation of
scales: We have \emph{hierarchical modularity}. Our goal is to measure 
quantitatively how this hierarchical modularity changes at different 
levels of random rearrangement.

\subsection{Random Rearrangement of the Vicsek Snowflake}
\label{themodel}
\begin{figure}[!t]
\begin{center}
\includegraphics*[height=180pt]{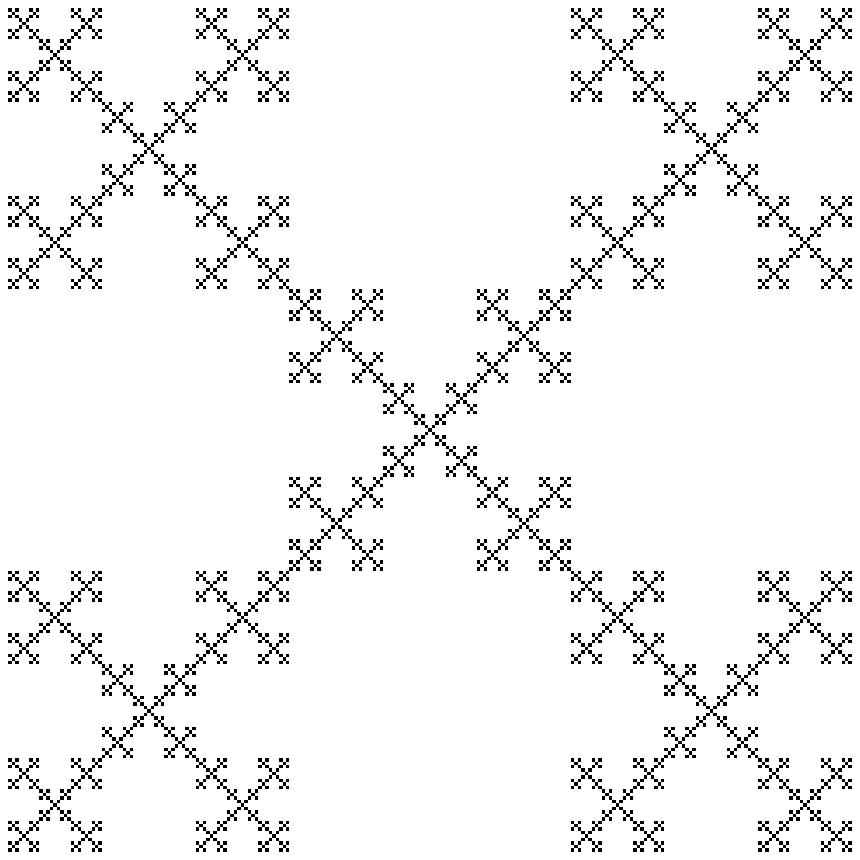}
\includegraphics*[height=180pt]{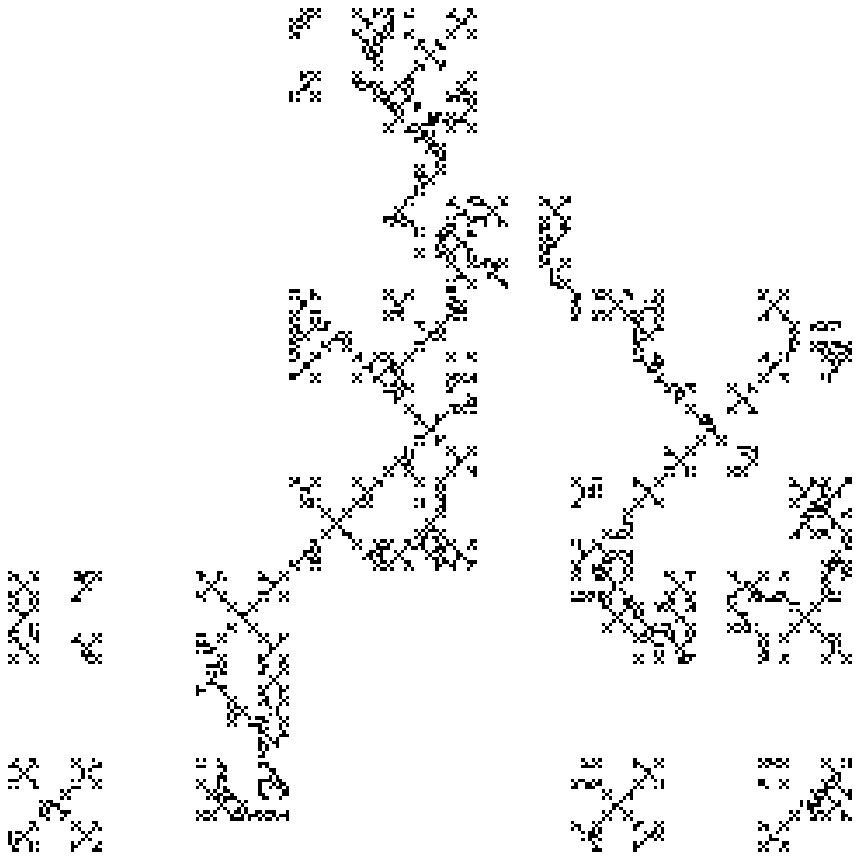}
\caption{Randomly rearranged fractals of size $N = 7$; $p = 0$ (left)
and $p = 0.2$ (right)}
\label{fig:otherps1}

\includegraphics*[height=180pt]{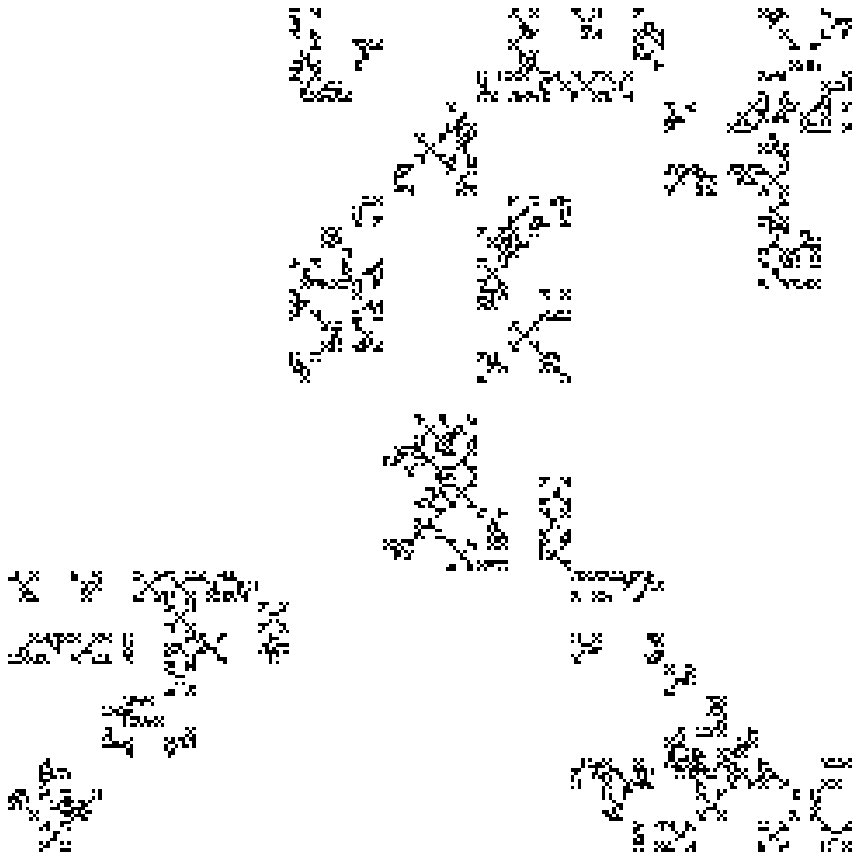}
\includegraphics*[width=180pt]{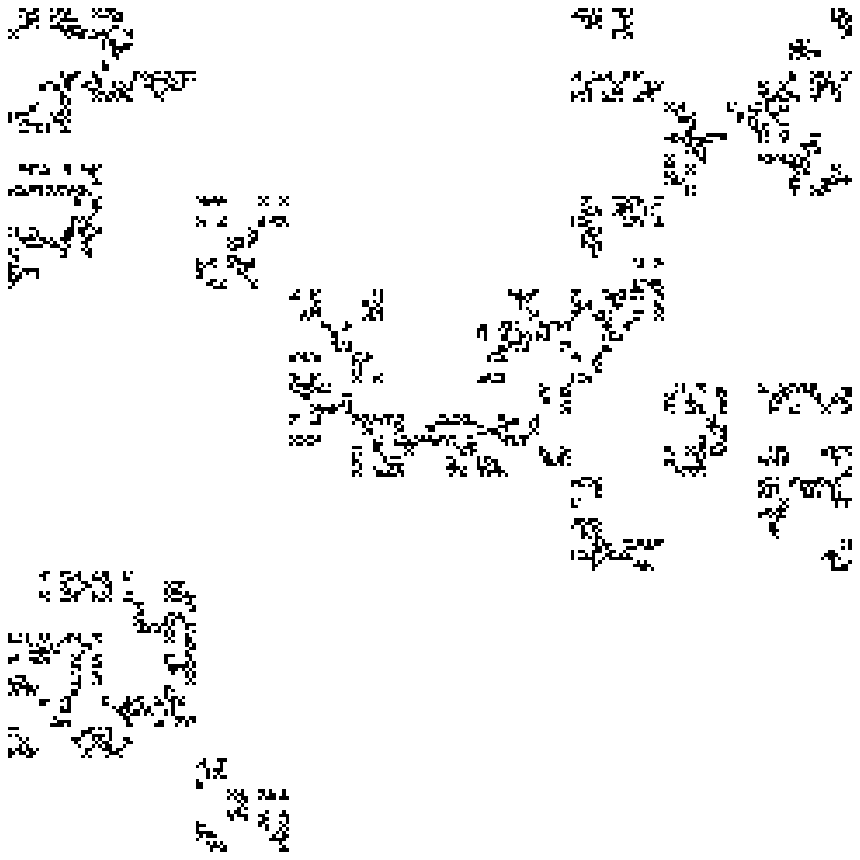}
\caption{Randomly rearranged fractals of size $N = 7$; $p = 0.4$ (left)
and $p = 0.6$ (right)}
\label{fig:otherps2}

\includegraphics*[width=180pt]{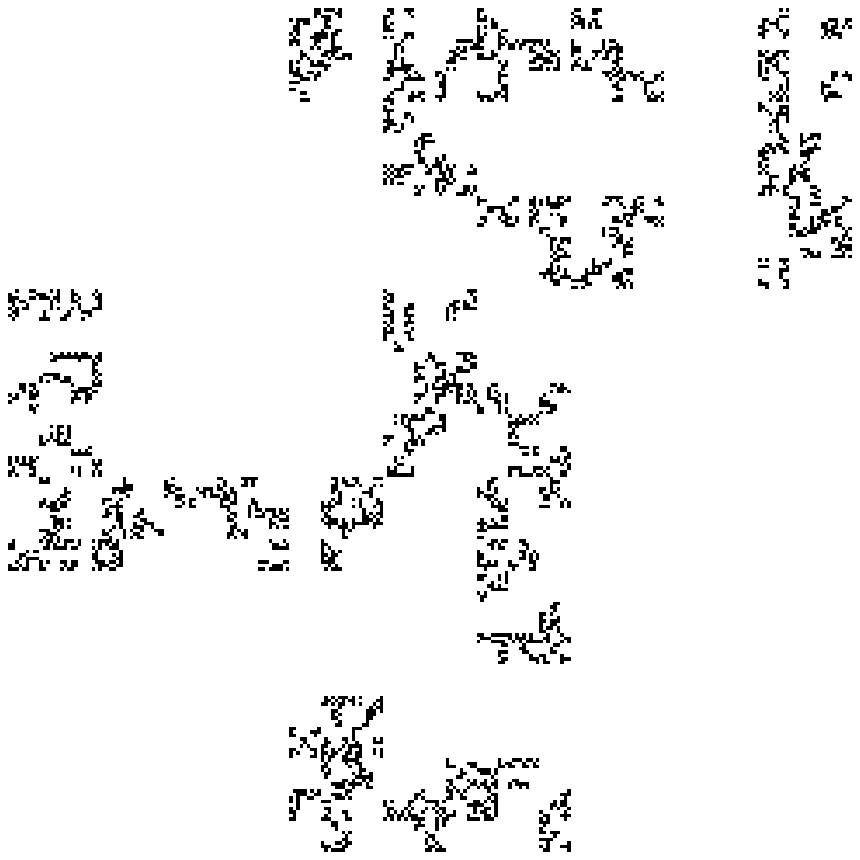}
\includegraphics*[width=180pt]{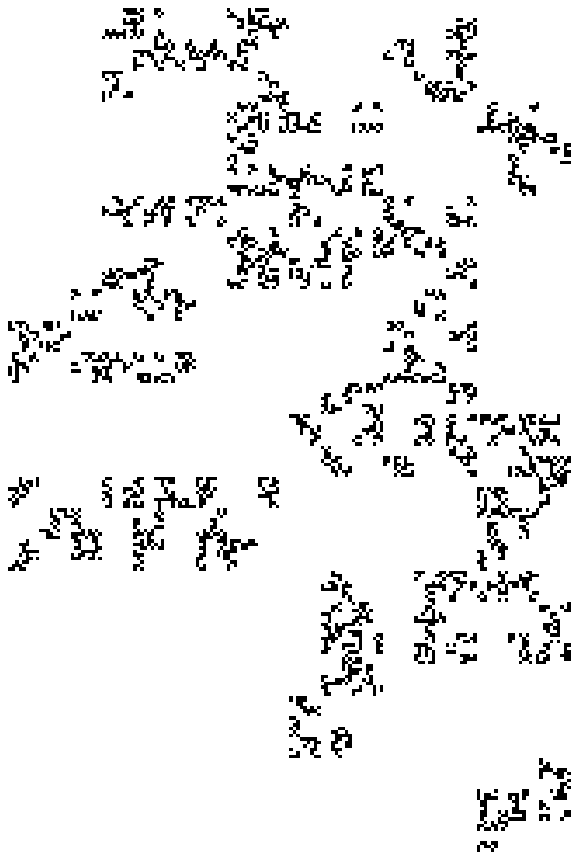}
\end{center}
\caption{Randomly rearranged fractals of size $N = 7$; $p = 0.8$ (left)
and $p = 1$ (right)}
\label{fig:otherps3}
\end{figure}

We generated the initial regular Vicsek snowflake by taking a $3^N
\times 3^N$ sized bit matrix evenly filled with \emph{ones}. One bit
corresponds to an elementary block of the fractal. Then, starting with
the highest hierarchical level, our algorithm 'cut out' the five
largest blocks, by turning the corresponding bits to
\emph{zero}. After that, taking the next sublevel, the next twenty
five sub-blocks were cut out from the five large blocks having been
generated in the previous step. After $N$ cutting steps one gets the
expected Vicsek snowflake with finite resolution. \newline The random
rearrangement algorithm is based on this cutting method above. We got a
continuous set of randomly rearranged fractals with a
randomness being controlled by the parameter $p\in[0,1]$ in the
following way: At each hierarchical level, we shifted all of the five
blocks with probability $p$ to the remaining free spaces in the larger
block. More precisely, we tossed for each of the five blocks whether
they would be shifted or not, then we cut out the non-shifted ones,
and \emph{after} that we randomly placed the remaining ones to the
free spaces with uniform distribution. So, before all of the cutting
steps our rearrangement algorithm made that draw described
above. \newline In this model $p = 0$ corresponds to the original
Vicsek snowflake, and $p = 1$ generates a totally random fractal, in
which all the five blocks take place in the nine rooms with uniform
random distribution at all hierarchical levels. By visual inspection
it is clear that the degree of hierarchical modularity decreases as
$p$ changes from $0$ to $1$ (see Figs. \ref{fig:otherps1} -
\ref{fig:otherps3}), and in case $p = 1$, it should be vanished (at
least on the average). Note that a $p = 1$ fractal is also made in a
hierarchical manner, but the applied random rearrangement destroys its
regular modular structure by mixing the modules together at
every hierarchical level.

\subsection{Measurement of Hierarchical Modularity}
\label{measurement}
In Ref. \cite{Vicsek} several methods for measuring the
dimension of regular fractals were compared. In order to quantify modularity,
we apply the \emph{box counting method}, where we cover growing
regions of the fractal with a growing square window started from the
centre, and at each step we count the mass (the number of elementary
blocks) inside the window. By this means, one gets a mass function:
the inside mass of the fractal as a function of the linear size of
the window. As pointed out in Ref. \cite{Vicsek}, for regular fractals this
mass function is not a straight line in a log-log plot: There are
periodic deviations due to the above mentioned separation of scales,
i.e., due to modularity. 
Therefore we consider the periodicity of these deviations from the
straight lines as the measure of modularity.
\newline From our model, presented in Sec.  \ref{themodel}, one gets 
different deviation functions for different values of $p$. In order to 
analyse these functions we applied Fast Fourier Transform (FFT) to
them.

\subsection{Results}
\label{results}
In this subsection we discuss the results of the measurement of
modularity described above in Sec. \ref{measurement}. 

\begin{figure}[!h]
\begin{center}
\includegraphics*[width=380pt]{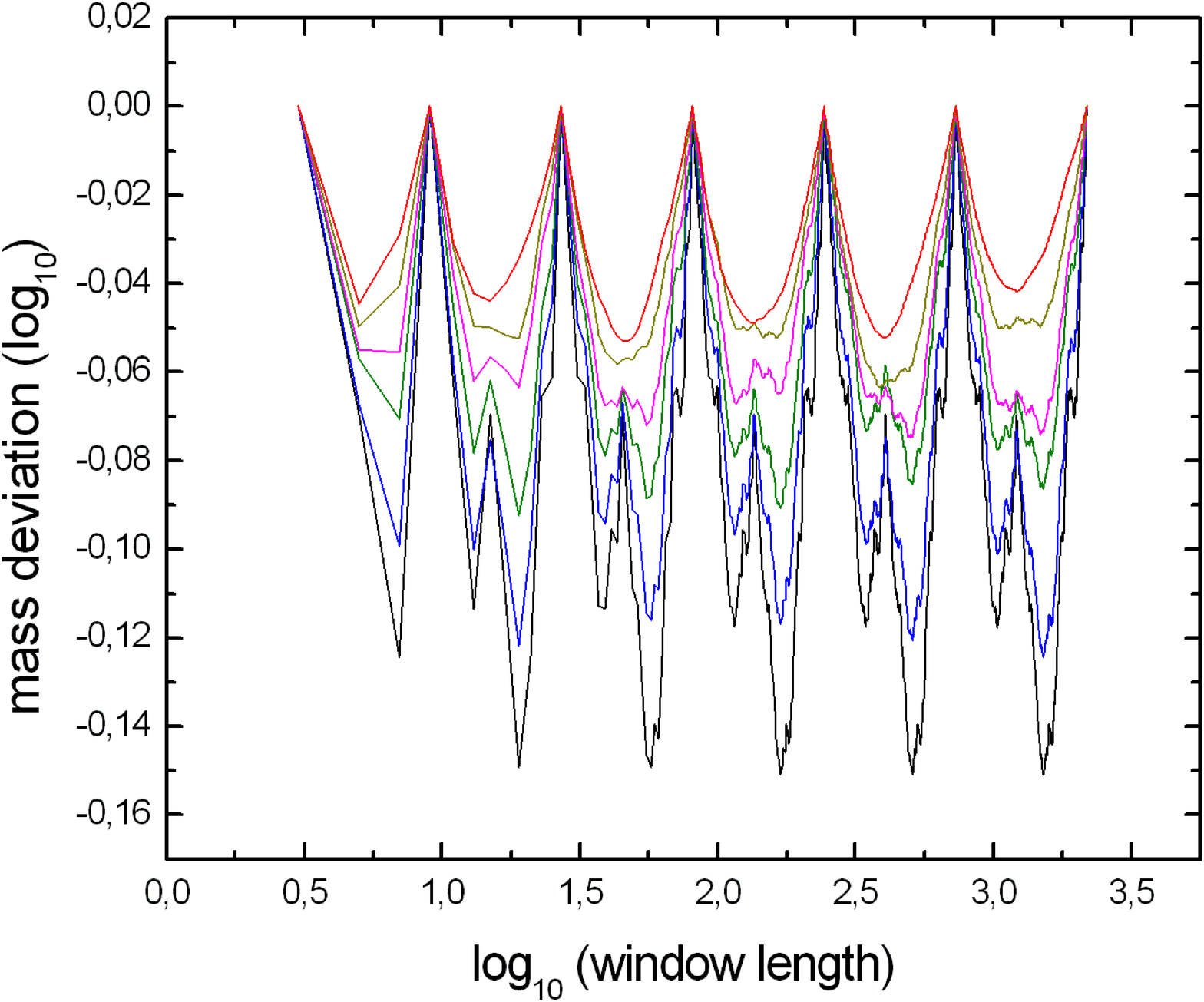}
\end{center}
\caption{The deviation functions for fractals $N = 7$ at different 
levels of randomization. The value of $p$ decreases by $0.2$ from above 
($p = 1$, red) to below ($p = 0$, black). In cases $p \neq 0$ we averaged 
200 randomly rearranged fractals.}
\label{fig:devfunc} 
\end{figure}

The log-log plots of the mass deviation functions as a function of the
window size are plotted in Fig. \ref{fig:devfunc} for different 
values of parameter $p$. For non-zero $p$, each curve is the average 
of $200$ randomly rearranged fractals (with same $p$).

\begin{figure}[!h]
\begin{center}
\includegraphics*[width=380pt]{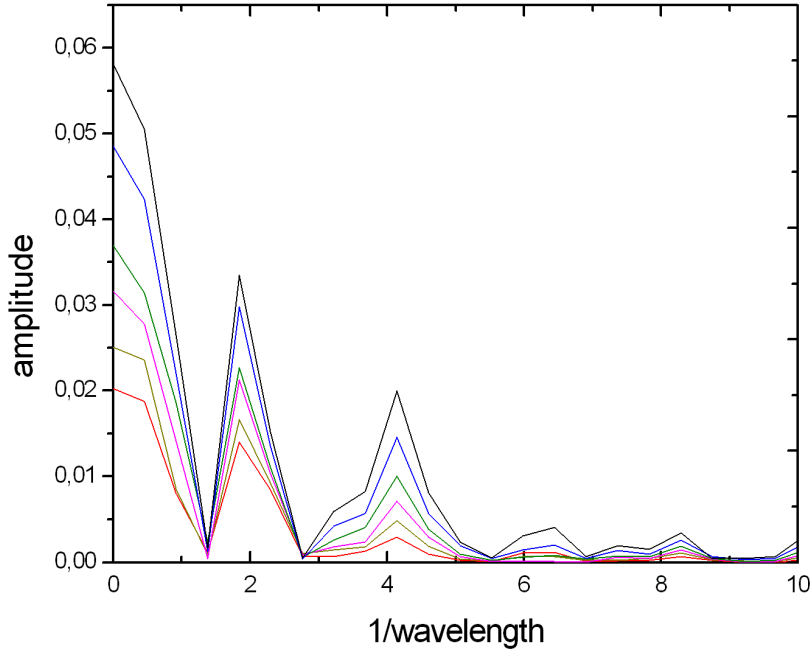}
\end{center}
\caption{The FFT spectrum of the deviation functions of Fig. \ref{fig:devfunc}.
The value of $p$ increases by $0.2$ from above ($p = 0$, black) to below
($p = 1$, red).}
\label{fig:FFT} 
\end{figure}

Apparently, the mass deviation functions reflect the hierarchical
modularity of the fractals of different $p$ values. In the $p = 0$
case, the function has an inherent structure corresponding to its high
degree of modularity. This scale-free structure is more visible on the
farther periods of the function, because there we have more
points. This inherent structure of the curves vanishes as
$p \rightarrow 1$. The maximum deviation also decreases as we increase
$p$, and the functions become smoother and smoother (apart from the
basic high peaks with values of $0$), indicating the vanishing
modularity of the fractals. The $0$-valued peaks correspond to powers
of $3$: when the linear size of the window reaches powers of $3$, the
window contains a whole sub-fractal with the exact dimension. This is
the consequence of not mixing the elements in a continuous way. The
deviation is non-positive for all values of $p$, which can be
explained by the geometry of the applied rectangular windowing method
in our special case. \newline The
vanishing modularity could be better represented taking the FFT
spectrum of the functions in Fig. \ref{fig:devfunc}. We analysed the
last three periods of the functions with FFT, where they have the
finest inherent structure. The results are plotted in
Fig. \ref{fig:FFT}.

Because of the triadic organization of the fractals, the fundamental
spatial frequency is $f = 1/\log 3 \approx 2.096$. This
corresponds to the $0$-valued peaks mentioned above, therefore the
first peaks in the FFT spectra are direct consequences of the
geometry of our method. For the regular fractal ($p = 0$) one could
notice the first, second and third harmonics which indicate the
scale-free property of the deviation function. The first harmonics,
which indicates the fine structure of the deviation functions,
significantly decreases for increasing values of $p$, and the second
and third harmonics totally disappear, what verifies the decreasing
degree of hierarchical modularity of the fractals for
increasing rearrangement probability $p$.

\section{MODULARITY OF NETWORKS}
\subsection{The Randomization of the Hierarchical Network Model}
\begin{figure}[!h]
\begin{center}
\includegraphics*[width=300pt]{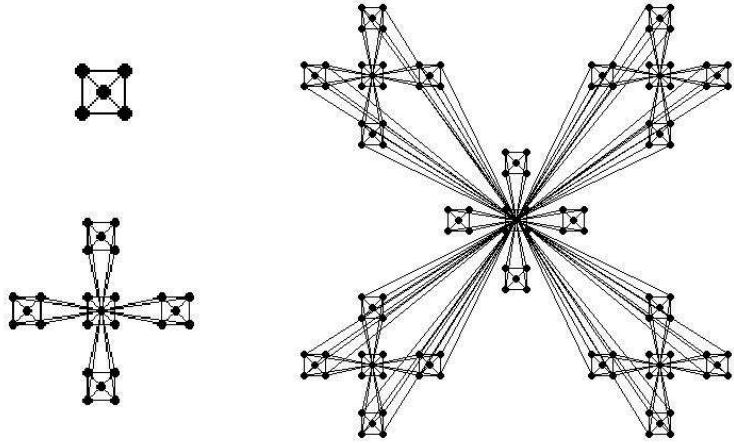}
\end{center}
\caption{The deterministic hierarchical network at different sizes \cite{RavaszBarabási}}
\label{fig:determ} 
\end{figure}

\begin{figure}[!h]
\begin{center}
\includegraphics*[width=300pt]{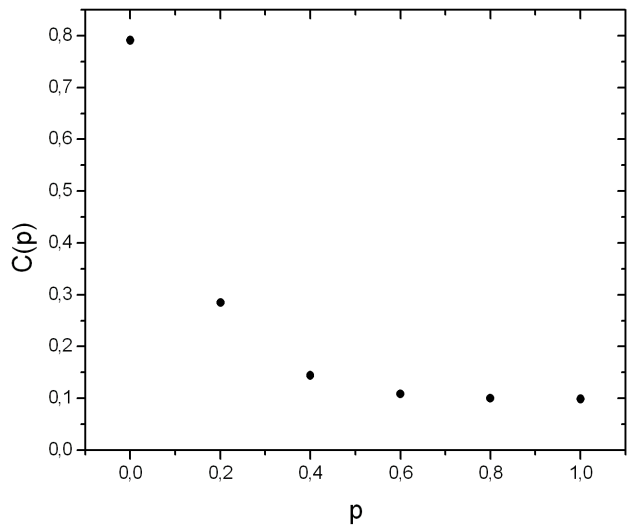}
\end{center}
\caption{The clustering coefficient as a function of the
randomization. Each point represents $200$ networks of $3125$ nodes.}
\label{fig:cp} 
\end{figure}

Our starting point was the deterministic, modular hierarchical network
model of Ravasz and Barabási \cite{RavaszBarabási} (see Fig. \ref{fig:determ}).
Its main features correspond to the real networks
\cite{RavaszBarabási}: the degree distribution is scale-free, the
clustering coefficient is independent of the size, and follows a
power-law as a function of degree: $C(k) \propto k^{-\alpha}$. In real
networks, $\alpha$ is usually about 1. 
\newline The regular model also has obviously 
hierarchically modular structure. As real grown networks are to some extent 
random, the question raises how the modularity is influenced by the
randomness. In order study this problem, we 
used a link randomization procedure, earlier already applied to
investigate the influence of randomness on synchronization
\cite{koreai}. In this model two links were chosen
randomly, and one node of both 
links was exchanged between the two links. This process was executed $M
\times p$ times, where $M$ is the number of the links and $p$ is the
control parameter of the randomization. This method conserves the
degree distribution, as it does not change the degree of any node. 
\newline As $p$ increases from $0$ to $1$, the average clustering coefficient
$\bar {C}$ first falls rapidly then becomes constant (see Fig. \ref{fig:cp}).

\begin{figure}[!h]
\begin{center}
\includegraphics*[width=300pt]{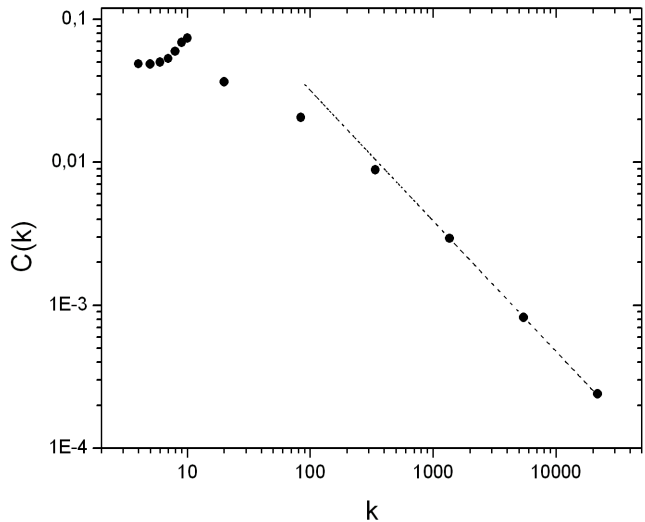}
\end{center}
\caption{Clustering as a function of degree. The slope of the dotted
line is $-0.913$. From the average of $400$ networks of $78125$ nodes,
$p = 1$.}
\label{fig:ck} 
\end{figure}

For $p = 1$, the degree dependence of $C(k)$ has two regions. It seems
that the very low degree ($k < 10$) nodes' behavior is significantly
different from the rest. Due to this effect and to the crossover it
causes, the asymptotics sets in rather late. However, it is clear
that the distribution is very broad and it has possibly a power law
tail (Fig. \ref{fig:ck}). Also due to the small $k$ anomaly the 
average $C$ decrease with increasing system size for the considered
number of nodes.

\subsection{Topological Overlap Matrix}
\begin{figure}[!h]
\begin{center}
\includegraphics*[width=400pt]{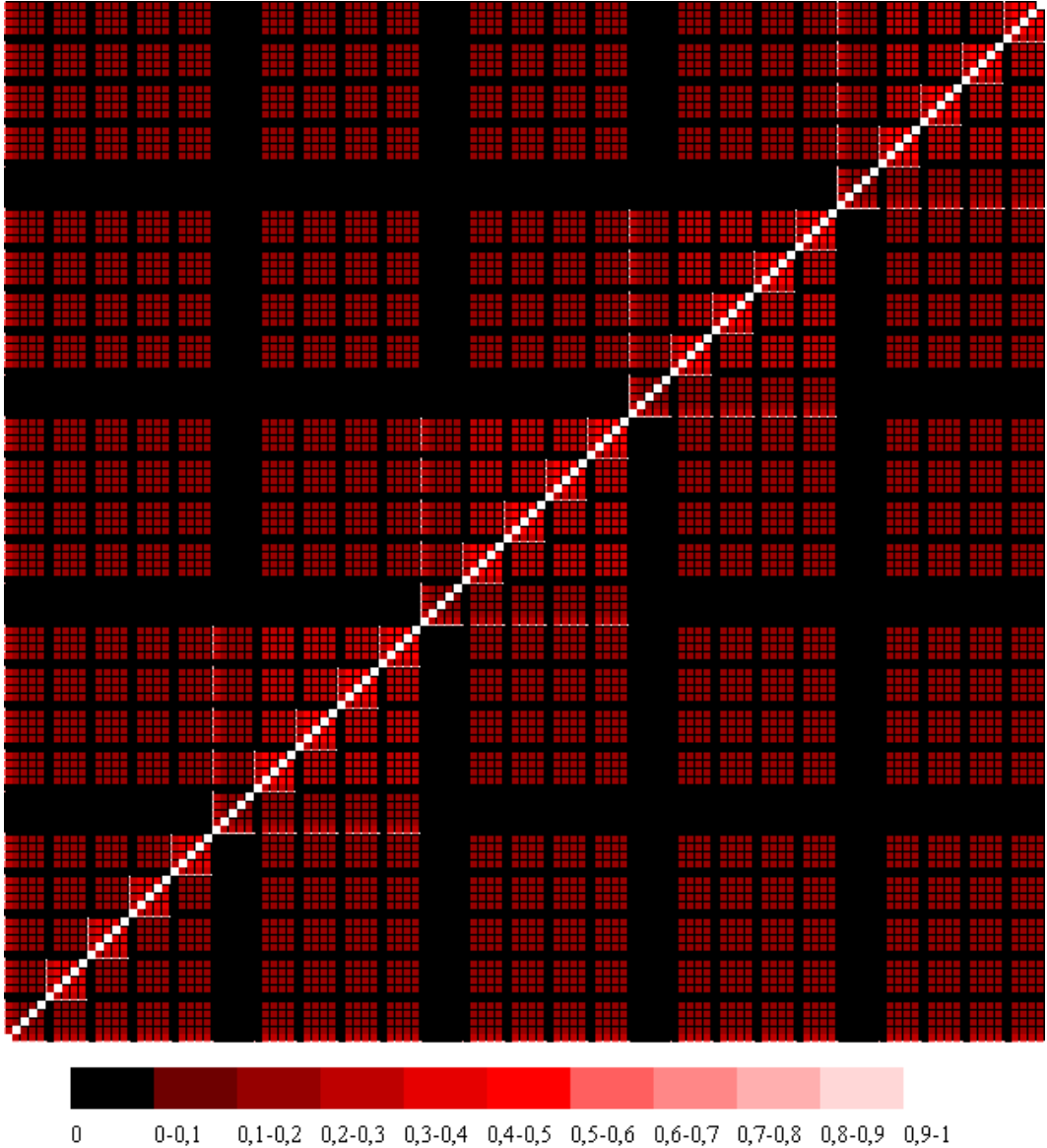}
\end{center}
\caption{Topological overlap matrix of the original deterministic
hierarchical network. The number of nodes is $625$.}
\label{fig:tom1} 
\end{figure}

To investigate the presence or absence of modules and hierarchical
structure, we used the topological overlap matrix (TOM) of the network
\cite{Ravasz}. The $ij$ element of this matrix, $T_{ij}$ equals the
number of mutual neighbours of the nodes $i$ and $j$ (plus $1$ if $i$
and $j$ are connected), normalized by the minimum degree of $i$ and
$j$, so $T_{ij}$ is between $0$ and $1$. Thus, the $i$-th row and 
column represents the overlap of the node $i$ with the other nodes of 
the network ($T_{ii}$ is defined as $1$). If there is an isolated 
module of densely interconnected nodes in the network, and 
rows/columns representing the nodes of the module are close to each 
other in the topological overlap matrix, the module appears in the 
matrix as a square centred on the diagonal with elements close to 
$1$. Of course, to enable this interpretation, the right sequence of 
the nodes has to be determined, just according to their overlap
values, as it is outlined in the next paragraph. The hierarchy of modules
is represented by a system of smaller and smaller (and more and more
cohesive) squares embedded into the larger squares (in which the
overlap of the modules decreases with the module size). These features
can be easily observed on the matrix of the deterministic model (Fig. \ref{fig:tom1}).

The sequence of the rows/columns representing the nodes in the matrix is
essential to recognize the modules and hierarchy. The 
rows/columns concerning nodes with big overlap have to be 
next to each other in the TOM, forming squares centred on the 
diagonal. In order to get the right sequence, we slightly modified the 
average linkage clustering method \cite{Ravasz,Eisen}. The original 
algorithm builds communities joining nodes into 'supernodes' (see Fig. ??: hierarchical tree, after a contraction the order in the new supernode has to be determined). In each step
two nodes are joined, meanwhile the TOM decreases by one row and column. 
\newline The basic steps of the algorithm:

\begin{itemize}
\item[-] First it finds the highest element in the TOM and contracts
the two corresponding nodes into a supernode. 
\item[-] The corresponding rows and columns of the two original nodes in the matrix are
contracted into 1 row and 1 column (matrix elements are averaged),
then the next round is started.
\item[-] This procedure is repeated until the TOM decreases into a $1\times1$ matrix 
(every node joined into one supernode).
\end{itemize}

The previous steps describe the original algorithm. Because it was unable to reconstruct 
even the simple regular case shown on Fig. \ref{fig:tom1}, some modifications were applied:

\begin{itemize}
\item[-] It is possible that the above algorithm finds more than one elements with the same high value in the same step. In this case the contraction resulting the smallest supernode is performed. If this quantity is also degenerated, then the choice is made at random. The goal of this modification is to make the growing of the supernodes
\item[-] We are searching for clusters, not just pairs. Therefore the algorithm examines that the selected contraction is good from the view of building a cluster (containing more than $2$ nodes). If a better contraction is possible, that will be executed.
\end{itemize}

This way a sequence of contractions emerges. Parallel with the above algorithm placing the contracted nodes next to each other, the 'clustered' sequence of the nodes appear. To make the above algorithm more clear, there is a small example:

\begin{figure}[!h]
\begin{center}
\includegraphics*[width=400pt]{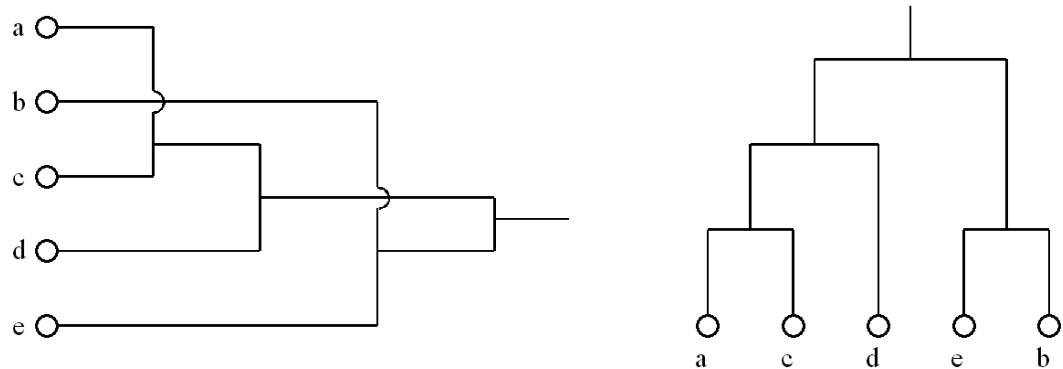}
\end{center}
\caption{On the left-hand side, the contraction process is shown for $6$ nodes. On the right-hand side, the resulting dendogram is displayed. Below: the corresponding matrices at timestep $t=0\textrm{(the original TOM)}, 1, 2, 3$. The grey characters are out of interest.}
\label{fig:kieg} 
\end{figure}

\begin{scriptsize}
\begin{tabular}{l l r}

\begin{tabular}{c|c c c c c}
0 & a & b & c & d & e \\
\hline
a & & $.1$ & $.7$ & $.5$ & $.3$ \\
b & & & $.2$ & $.1$ & $.3$ \\
c & & & & $.4$ & $.3$ \\
d & & & & & $.3$ \\
\end{tabular}
\phantom{a}

&

\begin{tabular}{c|c c c c c}
1 & ac & b & e & d & \textcolor[gray]{0.5}{e} \\
\hline
ac & & $.15$ & $.3$ & $.45$ & \textcolor[gray]{0.5}{$.3$} \\
b & & & $.3$ & $.1$ & \textcolor[gray]{0.5}{$.3$} \\
e & & & & $.3$ & \textcolor[gray]{0.5}{$.3$} \\
\textcolor[gray]{0.5}{d} & & & & & \textcolor[gray]{0.5}{$.3$} \\
\end{tabular}
\phantom{a}

&

\begin{tabular}{c|c c c c c}
2 & acd & b & e & \textcolor[gray]{0.5}{d} & \textcolor[gray]{0.5}{e} \\
\hline
acd & & $.13$ & $.3$ & \textcolor[gray]{0.5}{$.45$} & \textcolor[gray]{0.5}{$.3$} \\
b & & & $.3$ & \textcolor[gray]{0.5}{$.1$} & \textcolor[gray]{0.5}{$.3$} \\
\textcolor[gray]{0.5}{e} & & & & \textcolor[gray]{0.5}{$.3$} & \textcolor[gray]{0.5}{$.3$} \\
\textcolor[gray]{0.5}{d} & & & & & \textcolor[gray]{0.5}{$.3$} \\
\end{tabular}\\

\end{tabular}
\vskip0.1cm
\begin{tabular}{l}

\begin{tabular}{c|c c c c c}
3 & acd & be & \textcolor[gray]{0.5}{e} & \textcolor[gray]{0.5}{d} & \textcolor[gray]{0.5}{e} \\
\hline
acd & & $.215$ & \textcolor[gray]{0.5}{$.3$} & \textcolor[gray]{0.5}{$.45$} & \textcolor[gray]{0.5}{$.3$} \\
\textcolor[gray]{0.5}{b} & & & \textcolor[gray]{0.5}{$.3$} & \textcolor[gray]{0.5}{$.1$} & \textcolor[gray]{0.5}{$.3$} \\
\textcolor[gray]{0.5}{e} & & & & \textcolor[gray]{0.5}{$.3$} & \textcolor[gray]{0.5}{$.3$} \\
\textcolor[gray]{0.5}{d} & & & & & \textcolor[gray]{0.5}{$.3$} \\
\end{tabular}\\

\end{tabular}
\end{scriptsize}

\begin{figure}[!h]
\begin{center}
\includegraphics*[width=180pt]{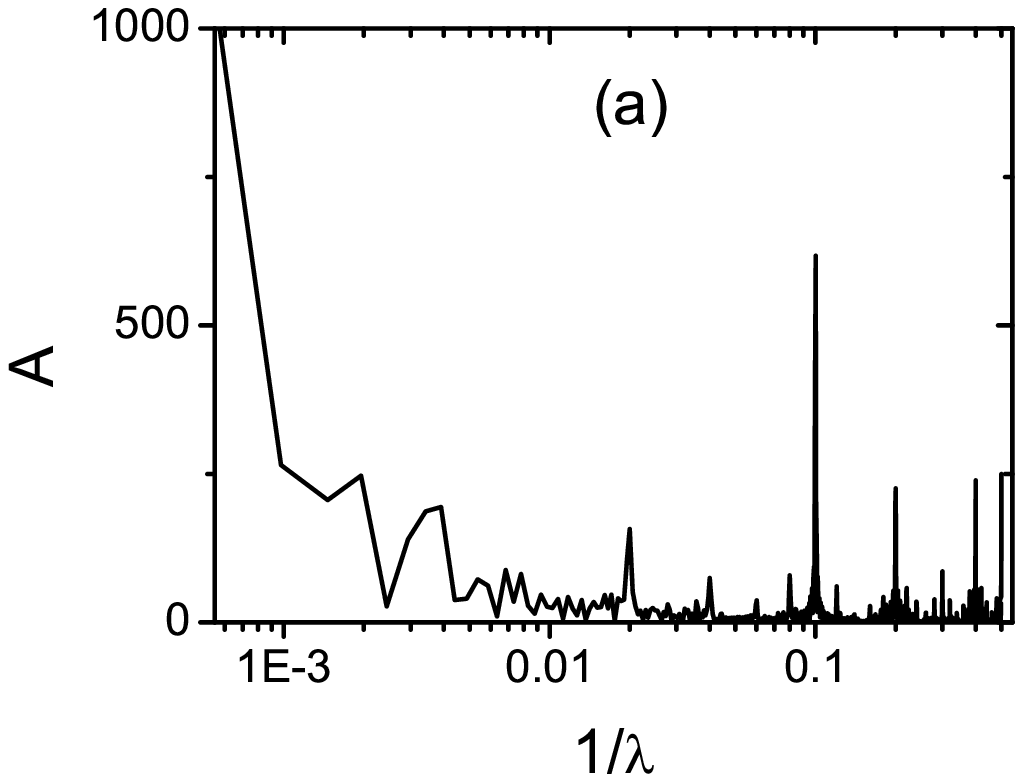}
\includegraphics*[width=180pt]{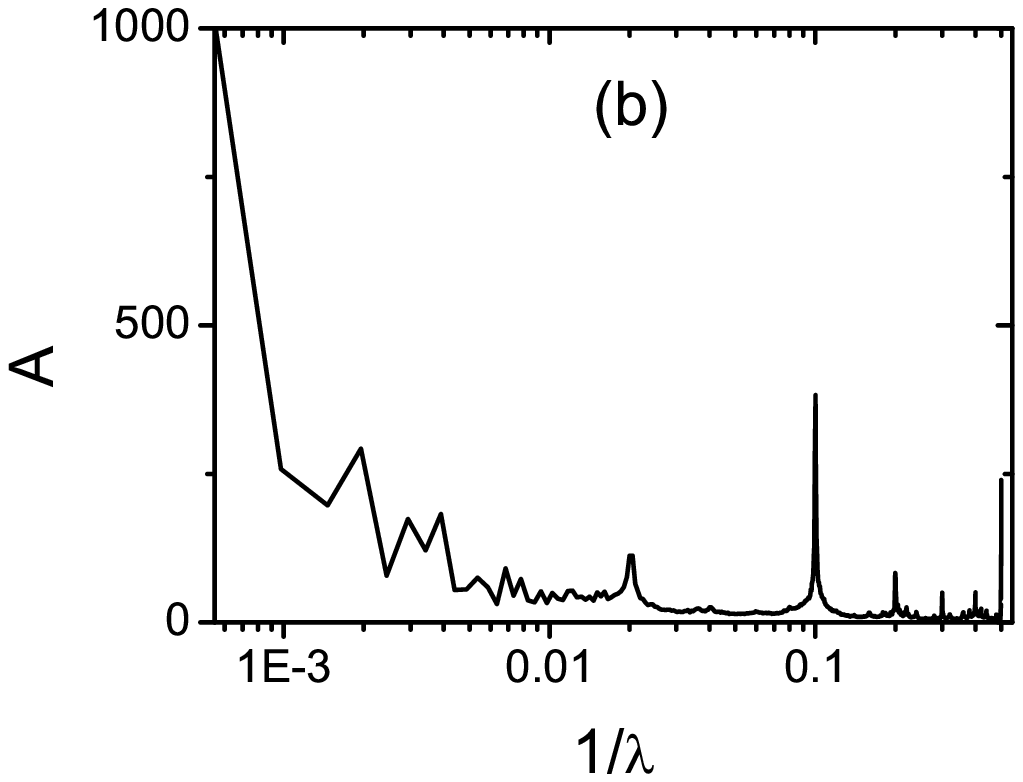}
\includegraphics*[width=180pt]{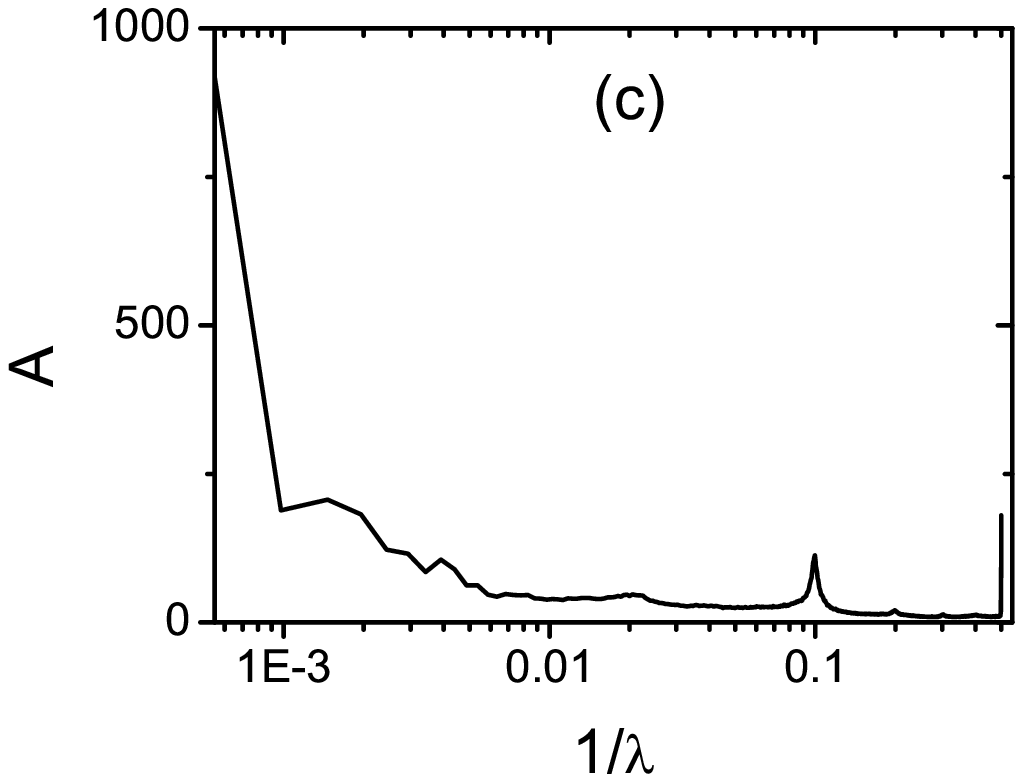}
\includegraphics*[width=180pt]{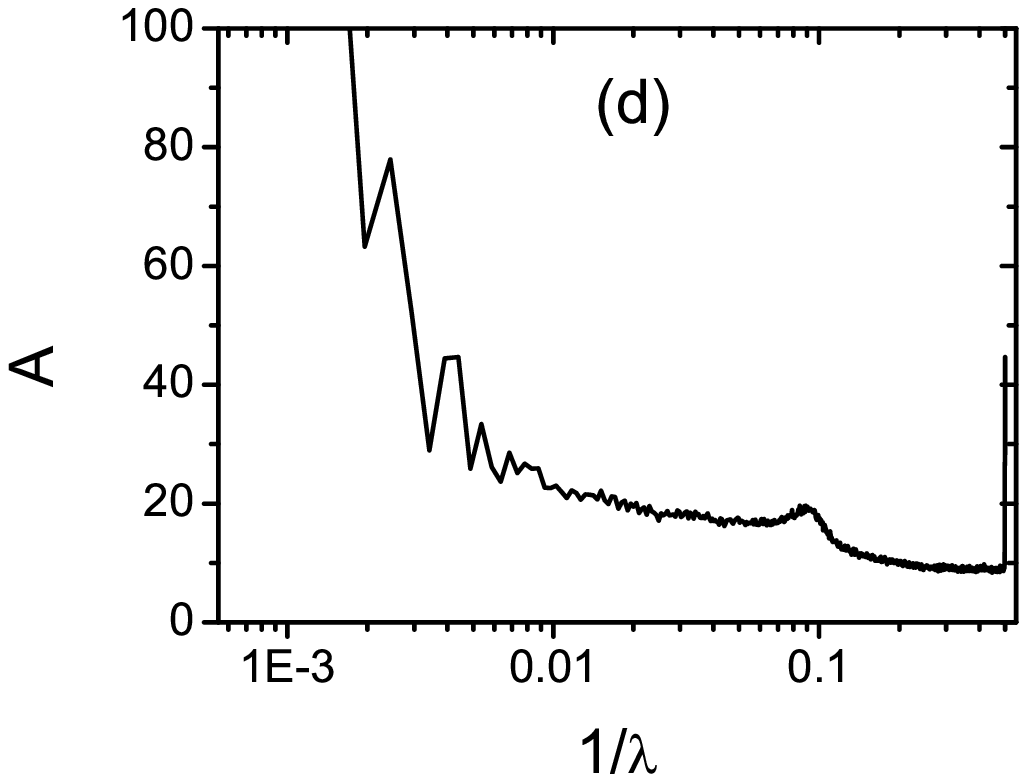}
\includegraphics*[width=180pt]{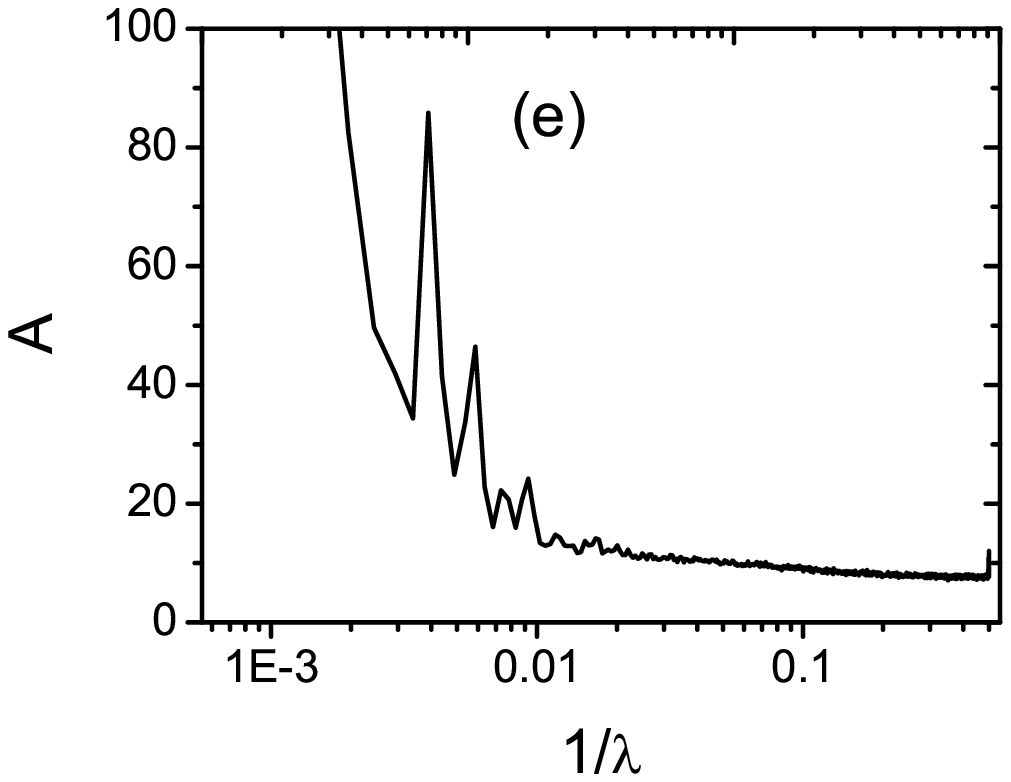}
\includegraphics*[width=180pt]{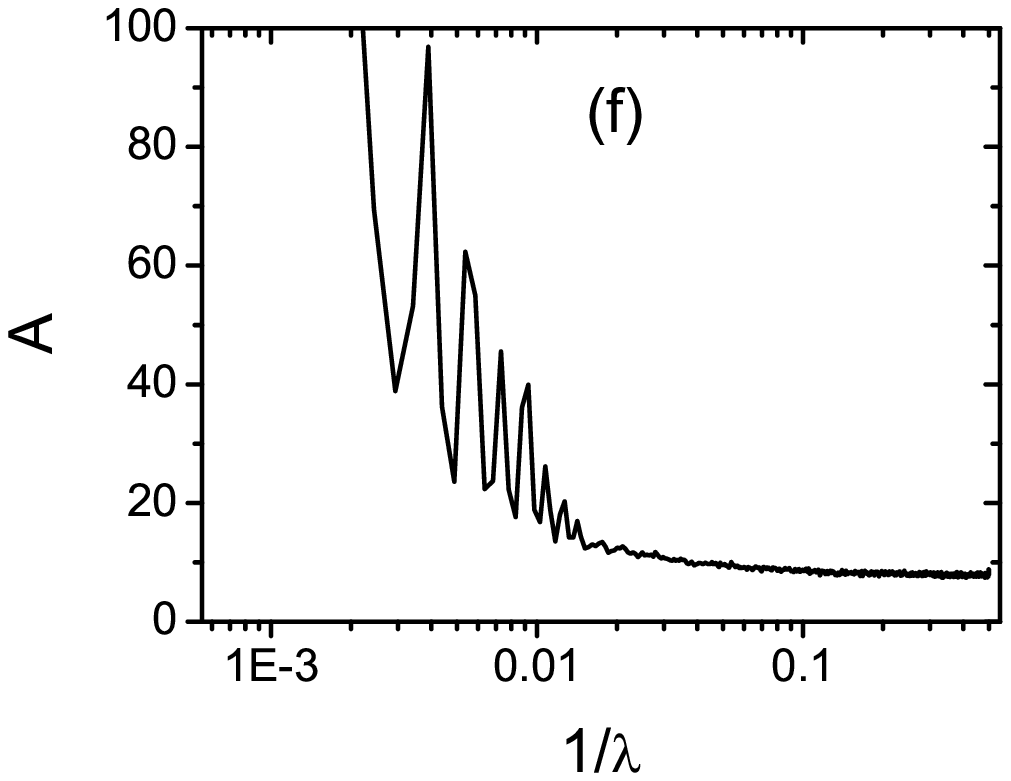}
\end{center}
\caption{FFT spectra of topological overlap matrices. Each curve
represents an averaged spectrum over 500 networks of 625 nodes for $p
= 0$ (a), $0.01$ (b), $0.05$ (c), $0.2$ (d), $0.4$ (e), $1$ (f).}
\label{fig:tomfft} 
\end{figure}

\begin{figure}[!h]
\begin{center}
\includegraphics*[width=300pt]{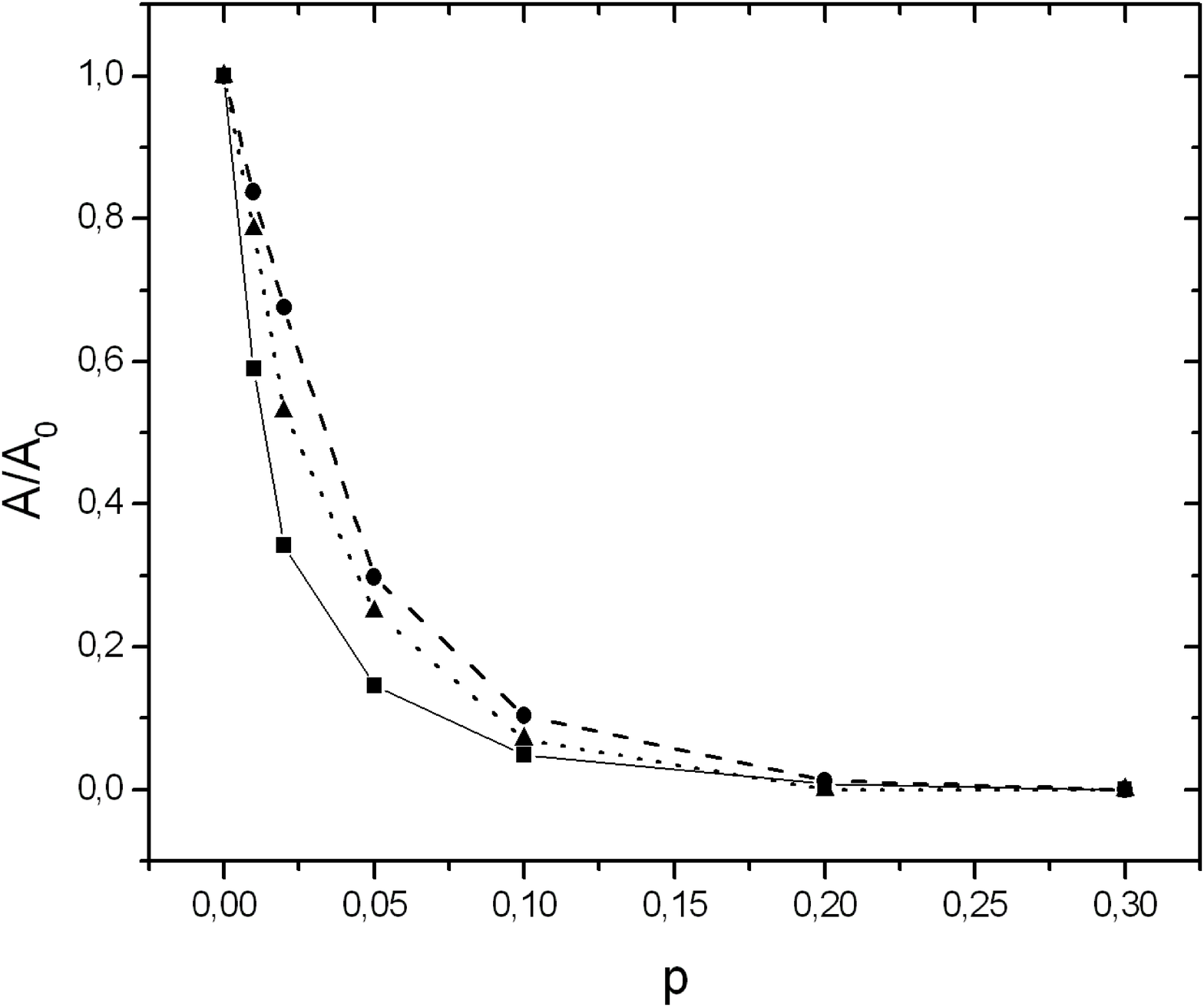}
\includegraphics*[width=300pt]{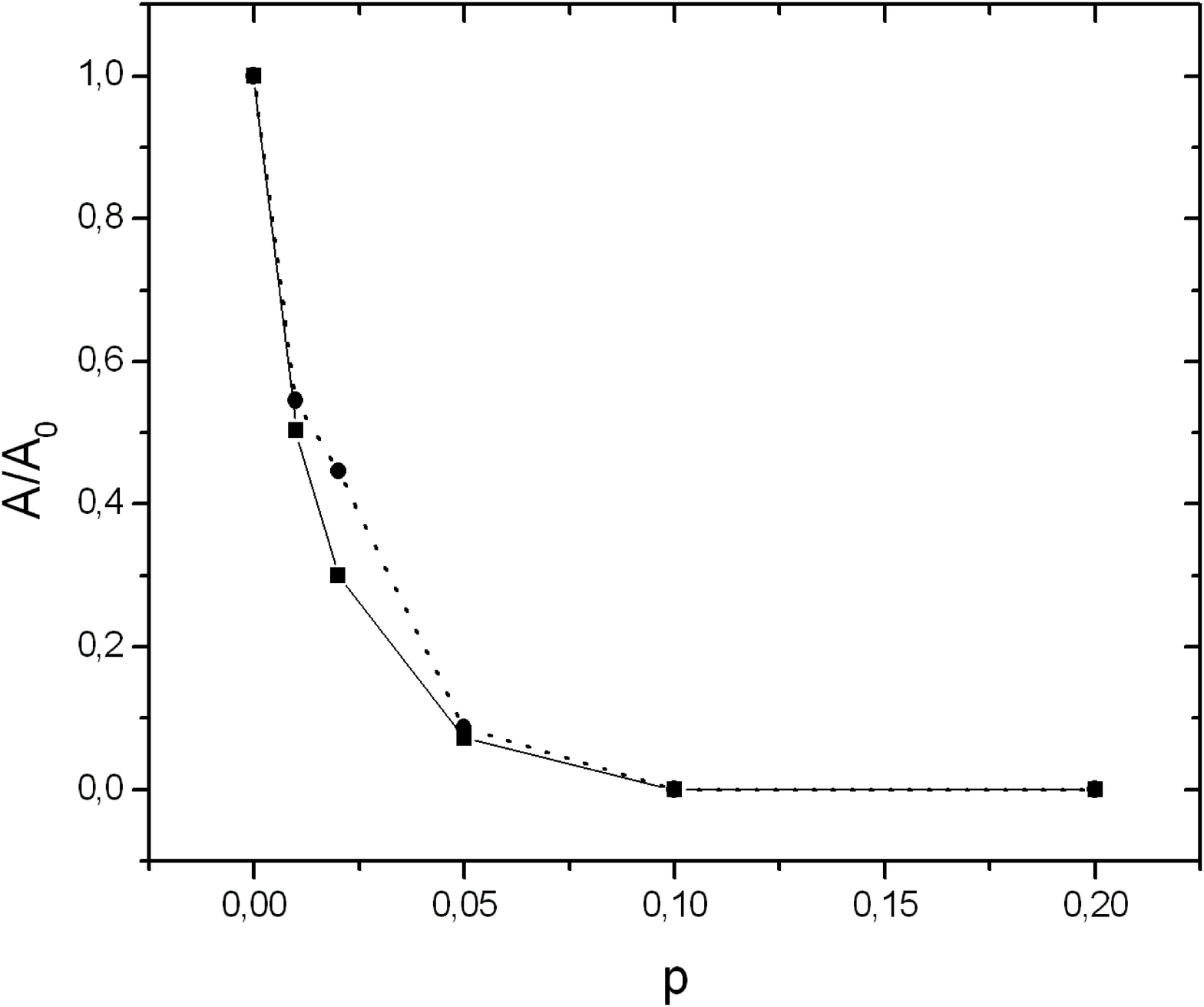}
\end{center}
\caption{FFT peaks corresponding to the $5$- (a) and $25$-node (b) clusters
of the deterministic, hierarchical network as functions of
randomization, averages over $500$ networks of size $625$ (squares),
$125$ (circles), $25$ (triangles). The amplitudes are normalized by the
value corresponding to $p = 0$. }
\label{fig:tomfftpeak} 
\end{figure}

Applying this algorithm, the TOM of the randomized deterministic
network is visually interpretable (Fig. \ref{fig:tom2}):
\begin{figure}[!h]
\begin{center}
\includegraphics*[width=400pt]{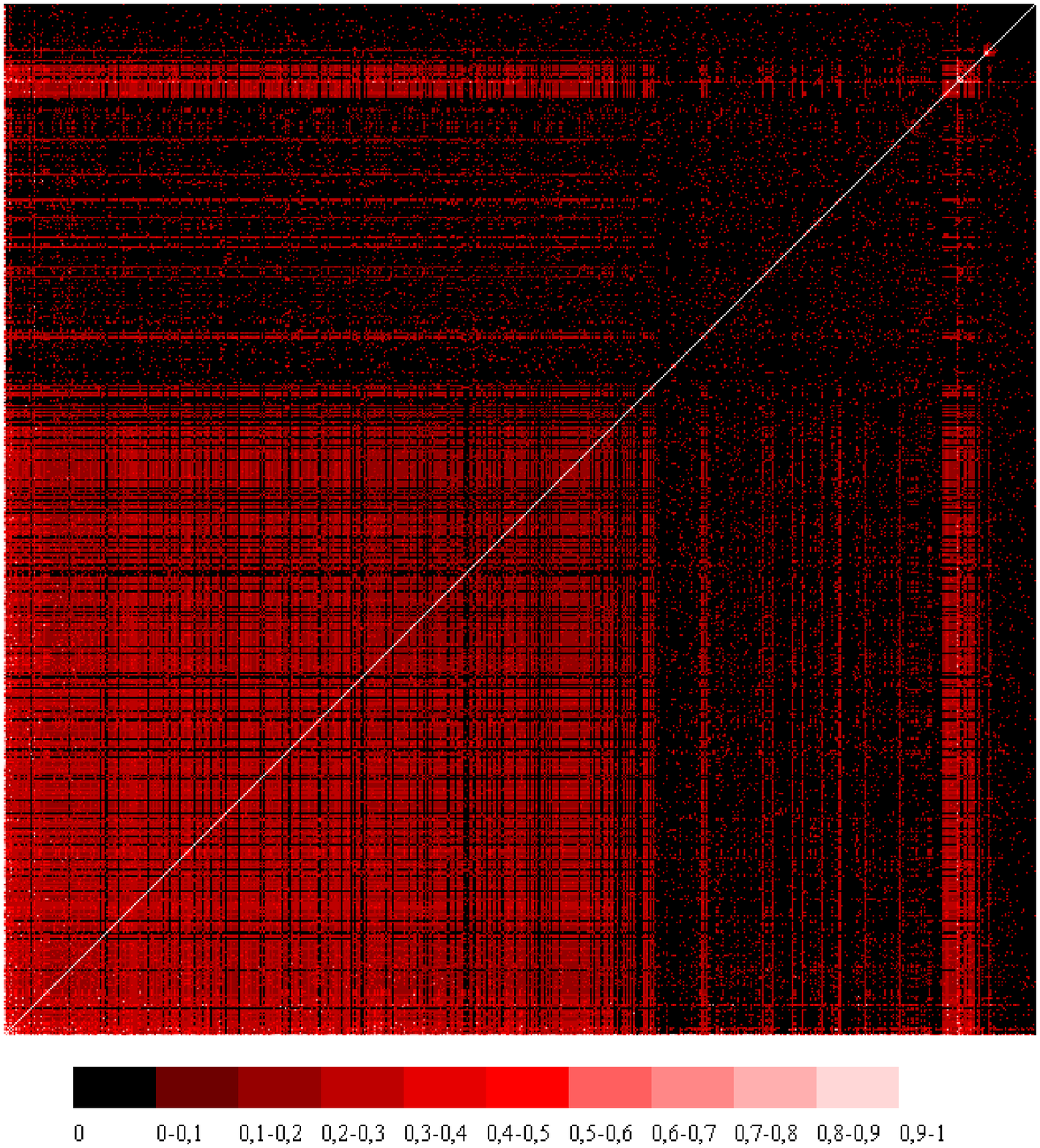}
\end{center}
\caption{The topological overlap matrix of the randomized ($p = 1$)
deterministic hierarchical network ($625$ nodes).}
\label{fig:tom2} 
\end{figure}
modular organization is not recognizable
with this clustering algorithm after the randomization process for $p=1$. 
This network is therefore scale free, it has a broad, probably power
law $C(k)$ distribution -- without a modular structure.
\newline We would like to make the vanishing of the regular modularity quantitatively
accessible in a similar way as in the case of the fractals: First, the
elements of the TOM are raised to the third power in order to
weaken the influence of the many small elements (to ´make
contrast´). Then the elements are projected (summed) perpendicular to
the diagonal (note that it means $2N-1$ sums for a $N \times N$
matrix), and the result is Fourier-analysed (Fig. \ref{fig:tomfft}).
A peak in the average of the amplitude-spectra indicates the presence
of equal-sized modules. As $p$ increases from zero, the peaks
indicating the presence of the 5-node and 25-node modules are
decreasing to zero rapidly (see Fig. \ref{fig:tomfftpeak}).

\section{DISCUSSION}
\label{concl}

In this paper we analysed the effect of randomness on the hierarchical
modularity of scale free structures. The quantitative analysis was
based on an important feature of modularity: the separation of scales.
We studied regular structures (Vicsek snowflake, deterministic
modular network) where randomness was introduced such that scale
freeness and other features (broad distribution of $C(k)$) were
maintained. Appropriate characteristics of modularity were chosen
using Fourier components of the mass deviation function (for fractals)
or of projections of the elements of the TOM (networks). In both cases
we observed a significant decrease of hierarchical modularity with
increasing randomness. For networks, we also observed a rapid fall for 
small values of $p$ in the clustering coefficient and in the Fourier 
peaks, suggesting a crossover similar to the phenomenon described in 
Ref. \cite{koreai}.    

It has to be emphasized that the applicability of the Fourier analysis
is based on the fact that the regular, hierarchical structures had a
regular scale separation. If the separation of scales is less regular, hierarchical
modularity can still exist, however, sufficient randomization would
destroy it in this case too. In many real networks signatures
of hierarchical modularity could be found \cite{RavaszBarabási} and the
same is true for some model networks like the Holme-Kim
\cite{Holme-Kim,Szabó} network \cite{private}. This indicates
that the level of irregularity in these networks is far from that reached by
randomization in our models. 

\section{ACKNOWLEDGEMENTS}
Thanks are due to T. Vicsek, E. Ravasz and
A.-L. Barabási for important discussions.

\end{document}